\documentclass[9pt, conference]{IEEEtran}
\IEEEoverridecommandlockouts

\usepackage{amsmath,amsfonts,graphicx}
\usepackage{amssymb,textcomp,mathtools}
\usepackage{bm,upgreek,algorithm,hyperref}
\usepackage{multirow,booktabs,hhline,array}
\usepackage{cite,url,makecell,setspace}
\usepackage{xcolor}
\pdfoutput=1

\renewcommand{\vec}[1]{\bm{\mathrm{#1}}}

\title{Apollo: Band-sequence Modeling for High-Quality Audio Restoration}

\author{\IEEEauthorblockN{Kai~Li$^{\spadesuit,\clubsuit,*}$, Yi Luo$^{\clubsuit,*}$\\}
\IEEEauthorblockA{$^\spadesuit$Department of Computer Science and Technology, Tsinghua University, Beijing, China \\
$^\clubsuit$Tencent AI Lab, Shenzhen, China \\
tsinghua.kaili@gmail.com, oulyluo@tencent.com}
\thanks{$*$ The work was done while Yi Luo was at Tencent AI Lab and Kai Li was an intern there.}
}

\begin{document}
\maketitle

\begin{abstract}
Audio restoration has become increasingly significant in modern society, not only due to the demand for high-quality auditory experiences enabled by advanced playback devices, but also because the growing capabilities of generative audio models necessitate high-fidelity audio. Typically, audio restoration is defined as a task of predicting undistorted audio from damaged input, often trained using a GAN framework to balance perception and distortion. Since audio degradation is primarily concentrated in mid- and high-frequency ranges, especially due to codecs, a key challenge lies in designing a generator capable of preserving low-frequency information while accurately reconstructing high-quality mid- and high-frequency content. Inspired by recent advancements in high-sample-rate music separation, speech enhancement, and audio codec models, we propose Apollo, a generative model designed for high-sample-rate audio restoration. Apollo employs an explicit frequency band split module to model the relationships between different frequency bands, allowing for more coherent and higher-quality restored audio. Evaluated on the MUSDB18-HQ and MoisesDB datasets, Apollo consistently outperforms existing SR-GAN models across various bit rates and music genres, particularly excelling in complex scenarios involving mixtures of multiple instruments and vocals. Apollo significantly improves music restoration quality while maintaining computational efficiency. The source code for Apollo is publicly available at \url{https://github.com/JusperLee/Apollo}.
\end{abstract}
\section{Introduction}
\label{sec:introduction}
Audio restoration has gained widespread application across various scenarios, ranging from music playback to real-time communication systems. For instance, in restoring vintage music, audio restoration methods effectively rejuvenate classic music pieces eroded by time or constrained by outdated equipment \cite{lattner2021stochastic,liu2021voicefixer,chen2023gesper}. Moreover, these methods are found to be extensively used in speech communication, particularly in telephone or internet calls, by repairing low-quality or distorted codec audio at the receiving end, thereby delivering a clearer and more natural auditory experience \cite{deng2020exploiting,dietz2002spectral,backstrom2017speech,li2024audio,chen2023mc,li2024safeear}. In music playback, audio restoration mitigates the degradation caused by compression, ensuring that users enjoy high-fidelity audio \cite{deng2020exploiting,lemercier2024diffusion,moliner2023solving}. For generative models, such as those used in music generation and speech synthesis, the audio quality is crucial, and restoration methods can enhance data quality, thus significantly improving model performance \cite{ji2020comprehensive,uhlich2024sound,zeng2024instructsing}. Robust audio restoration methods have become indispensable components of modern audio processing systems \cite{li2024enrollment}.

Audio restoration involves predicting high-quality, undistorted audio from degraded or compressed inputs. Current audio restoration technologies primarily focus on vocal recovery \cite{deng2020exploiting,dietz2002spectral,backstrom2017speech}. In traditional methods, a common technique is bandwidth extension \cite{dietz2002spectral,backstrom2017speech}, which aims to reconstruct lost high-frequency information and improve the perceptual quality of highly compressed audio signals. High-frequency spectral extension enhances encoding efficiency and proves crucial in low-bitrate scenarios \cite{larsen2005audio}. However, in some cases, bandwidth extension can introduce high-frequency artifacts that may degrade the overall audio signal quality.

\begin{figure*}[h]
	\small
	\centering
	\includegraphics[width=2.0\columnwidth]{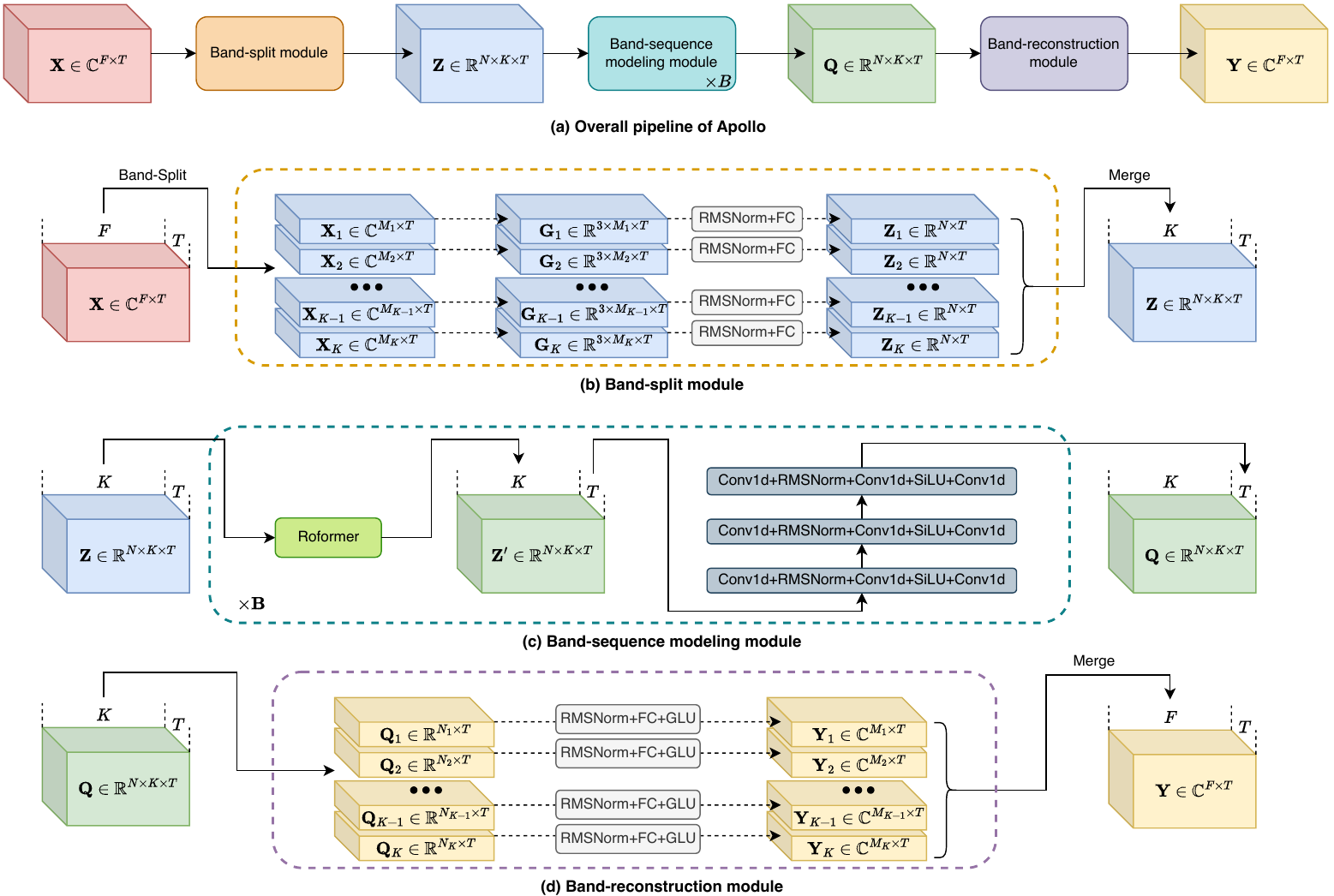}
	\caption{Overall pipeline of the model architecture of Apollo and its modules.}
	\label{fig:refiner}
\end{figure*}

With the rapid advancement of deep learning, NN-based methods have gradually replaced traditional signal-processing methods. Recently, GANs \cite{goodfellow2020generative} have demonstrated substantial potential in audio generation \cite{zeng2024joint}, super-resolution and restoration tasks \cite{lattner2021stochastic,pascual2017segan}, especially in achieving high-quality restoration. In audio codecs \cite{wu2023audiodec,kumar2024high,luo2024gull}, GANs effectively balance perceptual audio quality with distortion, offering superior restoration performance compared to traditional methods. Audio degradation typically affects the mid-to-high-frequency bands, particularly when using lossy codecs such as MP3 or AAC \cite{brandenburg1999mp3}, where high-frequency information is prone to compression artifacts. An ideal generator should retain the original audio's low-frequency components and supplement smooth and delicate mid-to-high-frequency details, thereby achieving a more realistic audio restoration effect. The Gull codec \cite{luo2024gull} has successfully demonstrated the effectiveness of GANs in the audio codec, showing significant progress in the super-resolution reconstruction of music and speech during the decoding phase of lossy codecs.

Inspired by Gull, we propose the Apollo model, a generative model specifically designed for high-sampling-rate audio restoration tasks. Apollo supports restoring audio quality at different compression rates. It comprises three main modules: a frequency band split module, a frequency band sequence modeling module, and a frequency band reconstruction module. Unlike Gull, we employ Roformer \cite{su2024roformer} in the frequency band sequence modeling module to capture frequency features and use TCN to model temporal features, enabling more efficient audio restoration. Specifically, Apollo first divides the spectrogram into sub-band spectrograms with predefined bandwidths, extracts gain-shape representations for each sub-band spectrogram, and encodes them through a bottleneck layer. Subsequently, stacked frequency band-sequence modeling modules perform interleaved modeling across frequency bands and sequences. Finally, each sub-band feature is mapped through nonlinear layers to generate the estimated restored sub-band spectrogram. These modules' design ensures the preservation of low-frequency information while restoring high-quality mid and high-frequency components. Additionally, with causal convolution and causal Roformer, our model supports streaming processing, making it suitable for real-time audio restoration.

We evaluated Apollo on the MUSDB18-HQ \cite{rafii2019musdb18} and MoisesDB \cite{pereira2023moisesdb} datasets, comparing it with state-of-the-art models such as SR-GAN \cite{lattner2021stochastic}. The experimental results showed that Apollo performed exceptionally well across various compression bitrates and music genres, particularly in complex scenarios involving a mixture of multiple instruments and vocals. Additionally, Apollo's efficiency in streaming audio applications has been validated, demonstrating its potential in real-time, high-quality audio restoration.

\section{Apollo}
\label{sec:model}
\subsection{Overall Pipeline}
\label{sec:overall}
Fig.\ref{fig:refiner}(a) presents the proposed Apollo pipeline. Apollo operates in the time-frequency domain and comprises a band-split module, a band-sequence modeling module, and a band-reconstruction module. Specifically, given compressed or distorted audio $\mathbf{S}\in \mathbb{R}^{1\times L}$, we first transfer $\mathbf{S}$ to its time-frequency domain representation $\mathbf{X}\in \mathbb{C}^{F\times T}$ using the Short-Time Fourier Transform (STFT), where $L$ denotes the length of audio, $F$ and $T$ denote the number of frequency bins and frames, respectively. Then, the band-split module maps to sub-band embeddings $\mathbf{Z}\in \mathbb{R}^{N\times K\times T}$ using gain-shape representations $\mathbf{G}\in \mathbb{R}^{3\times M\times T}$ for each sub-band, where $N$ and $M$ denote the number of channels in sub-band embeddings and gain-shape representations, respectively. Next, the band-sequence modeling module performs joint modeling of temporal and sub-band using a stacked architecture based on Roformer \cite{su2024roformer} and temporal convolutional network (TCN) \cite{bai2018empirical,li2022efficient}. Finally, the band-reconstruction module converts the output $\mathbf{Q}\in \mathbb{R}^{N\times K\times T}$ of the band-sequence modeling module into the reconstructed complex-valued spectrogram $\mathbf{Y}\in \mathbb{C}^{F\times T}$. It uses the inverse Short-Time Fourier Transform (iSTFT) to convert $\mathbf{Y}$ to a waveform $\bar{\mathbf{S}}\in \mathbb{R}^{1\times L}$.

\begin{table*}[]
\footnotesize
\centering
\caption{The structure of the STFT discriminator network.}
\begin{tabular}{cccccccc}
\toprule
\textbf{Layer Index} & \textbf{Layer Type}   & \textbf{Input Channels} & \textbf{Output Channels} & \textbf{Kernel Size} & \textbf{Padding} & \textbf{Stride} & \textbf{Activation} \\ \midrule
1                    & SpectralNorm + Conv2d & $F$                       & $F$                        & (3, 3)               & (1, 1)           & (1, 1)          & LeakyReLU(0.2)      \\
2                    & SpectralNorm + Conv2d & $F$                       & $F\times 2$              & (3, 3)               & (1, 1)           & (2, 2)          & LeakyReLU(0.2)      \\
3                    & SpectralNorm + Conv2d & $F\times 2$             & $F\times 4$              & (3, 3)               & (1, 1)           & (1, 1)          & LeakyReLU(0.2)      \\
4                    & SpectralNorm + Conv2d & $F\times 4$             & $F\times 8$              & (3, 3)               & (1, 1)           & (2, 2)          & LeakyReLU(0.2)      \\
5                    & SpectralNorm + Conv2d & $F\times 8$             & $F\times 16$             & (3, 3)               & (1, 1)           & (1, 1)          & LeakyReLU(0.2)      \\
6                    & SpectralNorm + Conv2d & $F\times 16$            & $F\times 32$             & (3, 3)               & (1, 1)           & (2, 2)          & LeakyReLU(0.2)      \\
7                    & Conv2d                & $F\times 32$            & 1                        & (3, 3)               & (1, 1)           & (1, 1)          & None               \\ \bottomrule
\end{tabular}
\label{tab:dis}
\vspace{-10pt}
\end{table*}

\subsection{Band-split Module}
As shown in Fig.\ref{fig:refiner}(b), given compressed or distorted audio spectrogram $\mathbf{X}$, we first split its frequency dimension $F$ into $K$ sub-band spectrograms $\{\mathbf{X}_k\in \mathbb{C}^{M_k\times T} | k\in [1, K]\}$. Inspired by the Gull codec \cite{luo2024gull}, we extract gain-shape representations $\mathbf{G}_k\in \mathbb{R}^{3\times M_k\times T}$ for each sub-band spectrogram:
\begin{equation}
\begin{aligned}
\mathbf{G}_{k} = \operatorname{Concat} \left[ 
\frac{\operatorname{Re}(\mathbf{X}_{k})}{\|\mathbf{X}_{k}\|_2}, \ 
\frac{\operatorname{Im}(\mathbf{X}_{k})}{\|\mathbf{X}_{k}\|_2}, \ 
\log\left(\|\mathbf{X}_{k}\|_2\right),
\right]
\end{aligned}
\end{equation}
where $\operatorname{Re}(\mathbf{X}_{k})$ and $\operatorname{Im}(\mathbf{X}_{k})$ denote the real and imaginary parts, respectively. $\|\mathbf{X}_{k}\|_2$ represents the $\ell_2$-norm of $\mathbf{X}_{k}$, given by:
\begin{equation}
    \|\mathbf{X}_{k}\|_2 = \sqrt{\operatorname{Re}(\mathbf{X}_{k})^2 + \operatorname{Im}(\mathbf{X}_{k})^2}
\end{equation}
$\log\left(\|\mathbf{X}_{k}\|_2\right)$ is the logarithm of the $\ell_2$-norm of $\mathbf{X}_{k}$. 
$\operatorname{Concat}$ refers to the concatenation of components. The gain-shape representation decouples the sub-band spectrogram's content and energy, allowing the reconstruction model to learn appropriate mappings that preserve the audio content. Subsequently, we map the gain-shape representations $\mathbf{G}$ into high-dimensional embeddings $\mathbf{Z}$ through a bottleneck layer, which consists of RMSNorm \cite{zhang2019root} and a 1D convolutional layer.

\subsection{Band-sequence Modeling Module}
In Apollo, we employ stacked Band-sequence modeling modules (BS modules, Fig.\ref{fig:refiner}(c)) to perform joint sub-band and temporal modeling with a stacking depth of $B$. Unlike BSRNN \cite{luo2023music} and Gull \cite{luo2024gull}, each BS module consists of a series of residual Roformers \cite{su2024roformer} and TCNs, which sequentially scan along the sub-band and time dimensions, and can increase the modeling capacity to improve the model performance. First, the residual Roformer is applied to the input $\mathbf{Z}$ along the frequency band dimension $K$ to obtain $\mathbf{Z}'\in \mathbb{R}^{N\times K\times T}$, capturing global dependencies between sub-bands while preserving the local characteristics of the frequency domain signals. Next, the TCN is applied along the time dimension $T$ on $\mathbf{Z}'$ to generate the output $\mathbf{Q}\in \mathbb{R}^{N\times K\times T}$. Since the $K$ sub-band features share the same feature dimension $N$, they all share a single TCN. The TCN consists of three convolutional blocks, each containing three convolutional layers. This design allows the TCN module to efficiently handle short-term dependencies and local temporal dynamics in audio signals, enhancing the model's ability to capture and understand temporal domain features.

\subsection{Band-reconstruction Module}
The output $\mathbf{Q}$ is passed through sub-band-specific fully connected (FC) layers to generate the estimated real and imaginary parts of the restored sub-band spectrograms (see Fig.\ref{fig:refiner}(d)). We utilize RMSNorm as the normalization layer within the fully connected layers and employ Gated Linear Units (GLUs) as the nonlinear activation function. Subsequently, the $K$ reconstructed sub-band spectrograms are concatenated along the frequency dimension to form the final reconstructed complex-valued spectrogram $\mathbf{Y}$. Finally, the reconstructed complex-valued spectrogram $\mathbf{Y}$ is converted back to the waveform domain $\bar{\mathbf{S}}$ through the iSTFT.

\begin{figure*}[h]
	\small
	\centering
	\includegraphics[width=2.0\columnwidth]{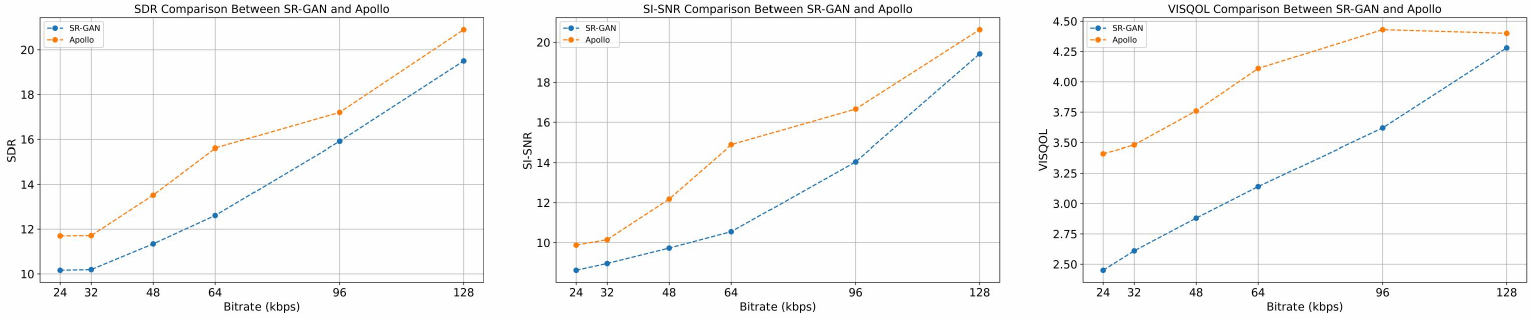}
	\caption{Apollo and SR-GAN's SDR, SI-SNR and ViSQOL result in comparison at different bitrates.}
	\label{fig:plot}
\end{figure*}

\begin{table*}[]
\footnotesize
\centering
\caption{Different methods' SDR/SI-SNR/VISQOL scores for various types of music, as well as the number of model parameters and GPU inference time. For the GPU inference time test, a music signal with a sampling rate of 44.1 kHz and a length of 1 second was used.}
\begin{tabular}{cccccccc}
\toprule
Model  & Vocal            & Single Stem      & Multi-Stems      & Multi-Stems+Vocal & Overall   & Params (M) & RTF (ms)      \\ \midrule
SR-GAN \cite{lattner2021stochastic} & 10.62/9.19/2.72  & 13.88/12.52/3.28 & 14.92/14.16/3.41 & 16.87/15.54/3.76  & 14.07/12.85/3.29 & 322.53 & \textbf{34.55}\\
Apollo (Ours) & \textbf{13.99/12.58/3.44} & \textbf{16.56/15.99/4.08} & \textbf{17.52/17.15/4.41} & \textbf{18.51/18.26/4.54}  & \textbf{16.64/16.00/4.12} & \textbf{16.54} & 53.23 \\ \bottomrule
\end{tabular}
\label{tab:stems}
\vspace{-10pt}
\end{table*}

\subsection{Training Objective}
The proposed Apollo model was trained using a GAN framework to enhance the quality of audio restoration. Specifically, the discriminator network is inspired by the multi-resolution STFT discriminator, similar to the Gull codec \cite{luo2024gull}. As described in Table \ref{tab:dis}, the discriminator input consists of real and imaginary parts of the spectrogram, which are stacked into a 3D tensor along the channel dimension. To ensure energy invariance in the input, the signal was normalized to have a unit $\ell_2$-norm before being passed into the discriminator. The discriminator is trained using the Least Squares GAN (LSGAN) loss \cite{mao2017least}, defined as:
\begin{equation}
    L_{\text{GAN}} = \sum_{i=1}^{I}\mathbb{E}_{\mathbf{A} \sim p_{\text{data}}} \left[ (D_i(\mathbf{A}) - 1)^2 \right] + \sum_{i=1}^{I}\mathbb{E}_{\mathbf{Y} \sim p_{\text{G}}} \left[ (D_i(\mathbf{Y}))^2 \right],
\end{equation}
where $\mathbf{A}\in \mathbb{C}^{F\times T}$ denotes the spectrogram of uncompressed audio and $I=5$ denotes the number of discriminator. 

The generator, Apollo, is optimized through a composite loss function, which includes the reconstruction loss, feature matching loss, and the adversarial loss from the discriminator. The \textit{reconstruction loss} $L_{\text{rec}}$ is based on the mean absolute error (MAE) between the magnitude spectrograms of the restored and target audio, evaluated over multiple STFT resolutions:
\begin{equation}
    L_{\text{rec}} =\frac{1}{W} \sum_{w=1}^{W} \frac{\left\| |\text{STFT}_{w}(\mathbf{Y})| - |\text{STFT}_{w}(\mathbf{A})| \right\|_1}{\left\| |\text{STFT}_{w}(\mathbf{A})| \right\|_1 },
\end{equation}
where $\text{STFT}_{w}$ denotes the STFT with window size $w\in [32, 64, 128, 256, 512, 1024, 2048]$. This multi-resolution approach allows the model to capture fine and coarse details, leading to accurate restoration of audio signals across various frequency ranges.

The \textit{feature matching loss} is defined as the layer-wise normalized MAE between the hidden representations of the discriminator for both the reconstructed and target signals. These hidden representations, denoted as $\bar{\mathbf{H}}_{i,j}$ for the reconstructed signal and $\mathbf{H}_{i,j}$ for the target signal, are obtained from the $j$-th layer of the $i$-th discriminator. The feature matching loss is computed as follows:
\begin{equation}
    L_{\text{FM}} = \frac{1}{5} \sum_{i=1}^{5}\left[\frac{1}{6} \sum_{j=1}^{6} \mathbb{E} \left[ \frac{\left| \bar{\mathbf{H}}_{i,j} - \text{sg}[\mathbf{H}_{i,j}] \right|}{\text{mean}\left( \left| \text{sg}[\mathbf{H}_{i,j}] \right| \right)} \right]\right].
\end{equation}
where $\text{sg}[\mathbf{H}_{i,j}]$ denotes $\mathbf{H}_{i,j}$ detached from the computational graph.

The overall generator loss combines reconstruction, feature matching, and adversarial losses, expressed as:
\begin{equation}
    L_{\text{G}} = \alpha L_{\text{rec}} + \beta L_{\text{FM}} + \gamma L_{\text{GAN}}
\end{equation}
where $\alpha=1$, $\beta=1$, and $\gamma=1$ are hyperparameters used to balance the contributions of the individual loss components. This comprehensive loss formulation ensures that Apollo reconstructs not only accurate audio signals but also maintains perceptual quality and adversarial robustness by leveraging multi-resolution STFT losses and feature-matching mechanisms.

\section{Experiment configurations}
\label{sec:config}
\subsection{Datasets}
We trained and tested Apollo on the combined MUSDB18-HQ \cite{rafii2019musdb18} and MoisesDB \cite{pereira2023moisesdb} datasets. By integrating these two datasets, we leveraged their rich diversity and comprehensive musical resources to evaluate Apollo's restoration performance across different music genres more thoroughly. During the data preprocessing stage, inspired by music separation techniques \cite{uhlich2024sound,li2024subnetwork}, we employed a Source Activity Detector (SAD) to remove silent regions from the tracks, retaining only the significant portions for training. Throughout the training, real-time data augmentation was implemented by randomly mixing tracks from different songs. Specifically, we randomly selected between 1 and 8 stems from 11 available tracks and extracted 3-second clips from each selected stem. These clips were then randomly scaled in energy within a range of [-10, 10] dB relative to their original levels. All selected stem clips were summed to generate simulated music. Subsequently, we simulated dynamic bitrate scenarios by applying MP3 codecs\footnote{\url{https://trac.ffmpeg.org/wiki/Encode/MP3}} with bitrates of [24, 32, 48, 64, 96, 128] kbit/s to generate the compressed music. To ensure that all samples were on the same scale, we rescaled both the target audio and encoded audio based on the maximum absolute value.

\subsection{Hyperparameters}
For the proposed Apollo model, the Short-Time Fourier Transform (STFT) window length was set to 20 ms with a hop size of 10 ms, using a Hanning window. The bandwidth for frequency band segmentation was set to 160 Hz, and the feature dimension $N$ was set to 256. The Band Sequence modeling module was stacked $B = 6$ times. In the discriminator network, the STFT window sizes were configured with a multi-scale setup, including $[32, 64, 128, 256, 512, 1024, 2048]$. For the optimizer, both the generator and discriminator utilize the AdamW optimizer \cite{loshchilov2017decoupled}. The generator's initial learning rate was set to 0.001, with a weight decay of 0.01, while the discriminator's initial learning rate was set to 0.0001, with the same weight decay of 0.01. The learning rate decayed by 0.98 every two epochs, and gradient clipping with a maximum norm of 5 was employed to prevent gradient explosion. Additionally, we implemented an early stopping mechanism to prevent overfitting: training was terminated if the validation loss did not decrease for 20 consecutive epochs. All the models were trained on eight NVIDIA RTX 4090 GPUs.

\subsection{Evaluation metrics}
In all experiments, we used the Scale-Invariant Signal-to-Noise Ratio (SI-SNR) \cite{le2019sdr}, Signal-to-Distortion Ratio (SDR) \cite{vincent2006performance}, and Virtual Speech Quality Objective Listener (VISQOL) \cite{hines2015visqol} to evaluate the quality of the reconstructed audio. To assess the model's efficiency, we reported the time consumption per second of audio processed by Apollo and SR-GAN (Real-Time Factor, RTF). RTF is calculated by processing 1-second audio tracks sampled at 44.1 kHz on both CPU and GPU, and the average value is taken after running 1000 iterations. Additionally, we measured the model size by reporting the number of parameters using the open-source tool PyTorch-OpCounter\footnote{\url{https://github.com/Lyken17/pytorch-OpCounter}}.

\section{Results}
\label{sec:result}
Due to the lack of openly available baselines for this task, it is not easy for us to make a fair comparison 
with other related works. We evaluated the restoration performance of the Stochastic-Restoration-GAN (SR-GAN) \cite{lattner2021stochastic} and Apollo models across various bitrates and music genres on the combined test set of MUSDB18-HQ and MoisesDB (with 5000 samples for each case). The test set encompasses a wide range of music genres, including vocals, single instruments, and mixed instruments, aiming to comprehensively assess each model's restoration capabilities.

\textbf{Bitrate Impact Analysis.} Fig.\ref{fig:plot} compares the performance of the Apollo model and the Stochastic-Restoration-GAN (SR-GAN) at different bitrates (ranging from 24 kbit/s to 128 kbit/s). The experimental results demonstrated that Apollo consistently outperformed SR-GAN across all bitrates, particularly in addressing issues such as frequency band voids or reduced signal bandwidth, as reflected by SI-SNR and SDR scores. Additionally, Apollo significantly improved audio restoration quality as measured by VISQOL. Project page\footnote{\url{https://cslikai.cn/Apollo/}} for Apollo's reconstructed audio given multiple MP3 bitrates.

\textbf{Music Genre Impact Analysis.} Table~\ref{tab:stems} further illustrates the performance of both models across different music genres. In audio scenarios involving vocals, single instruments, mixed instruments, and a combination of instruments with vocals, Apollo consistently surpasses SR-GAN, with its advantage being especially pronounced in complex scenarios with mixed instruments and vocals. This is attributed to Apollo's alternating band and sequence modeling design, which emphasizes capturing and restoring complex spectral information. Compared to SR-GAN, Apollo delivers higher user ratings (VISQOL) with comparable inference speed while maintaining a more compact model size. This is especially important for real-time communications and live audio restoration, where low latency is critical to the user experience.

\section{Conclusion}
\label{sec:conclusion}
We propose Apollo, a novel method specifically designed for compressed audio restoration. Apollo significantly enhances audio quality in the frequency domain through band split, sequence modeling, and reconstruction modules. Empirical evaluations on the integrated MUSDB18-HQ and MoisesDB datasets validate Apollo's outstanding performance. Notably, Apollo achieves substantial improvements in music restoration while maintaining a smaller model size and high computational efficiency. The experimental results demonstrated that when addressing the complex acoustic characteristics of music, band-split, and band-sequence modeling more effectively captured and restored audio information lost during compression.

\bibliographystyle{IEEEtran}
\bibliography{refs}

\end{document}